\documentclass[
aps
,prl
,twocolumn
,amsfonts,amssymb,amsmath
,twoside
,letterpaper
,bibnotes
]{revtex4}

\pdfoutput=1

\usepackage[pdftex]{graphicx}
\usepackage{sidecap}

\usepackage{bm,color,mathrsfs,hyperref}

\hypersetup{
    colorlinks=true, 
    linkcolor=blue,  
    citecolor=blue,  
    urlcolor=blue
}

\begin{document}

\title{
High impact of spin-nematic order on the lattice domains \\
in thin films of iron-based superconductors
}

\author{A. Cano}
\email{cano@esrf.fr}
\affiliation{
European Synchrotron Radiation Facility, 6 rue Jules Horowitz, BP 220, 38043 Grenoble, France}

\date{\today}

\begin{abstract}
The spin-nematic scenario proposed for the structural instability of iron-based superconductors is studied theoretically within a Ginzburg-Landau approach in epitaxial thin films. The transition temperature and the domains that appear due the interface with the substrate are computed analytically by following a variatonal procedure. It is predicted a cross-over thickness $l_*$ that separates the Kittel regime, in which the size of the domains goes as the square root of the film thickness $l$, from a new ultrathin regime in which the system tends to reach the single-domain state as $l \to 0$. The experimental observation of this cross-over is therefore proposed as a direct probe of spin nematiciy.
\end{abstract}

\pacs{
74.70.Xa, 
74.78.-w  
74.90.+n, 
}

\maketitle

\begin{figure}[b]
\centering\includegraphics[width=.475\textwidth]{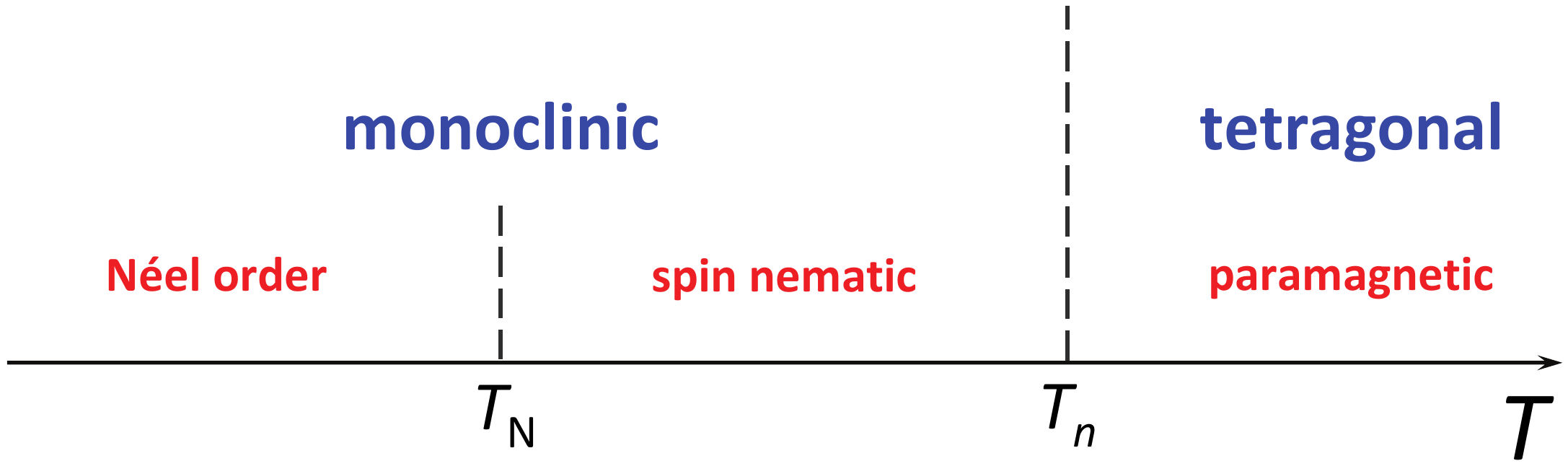}
\caption{Schematic representation of the spin-nematic scenario postulated for the iron-based superconductors. In the intermediate (spin-nematic) phase, there is a long-range correlation between the spin fluctuations in the Fe sublattices 
that breaks the tetragonal symmetry. The establishment of this spin-nematic phase from the paramagnetic state at $T_n$ is assumed to be behind the tetragonal $\leftrightarrow $ monoclinic structural transition in these systems.}
\label{scenario}
\end{figure}

The discovery of superconductivity in ferropnictides has triggered a huge interest on the properties of these materials \cite{reviews}. It is believed that magnetism plays a nontrivial role not only in the superconducting properties, but also in the structural features of these systems. Specifically, it has been suggested that the lattice distortion that precedes their antiferromagnetic transition is driven by a spin-nematic ordering of the electronic subsystem \cite{fang08} (see Fig. \ref{scenario}). At the conceptual level, the establishment of spin nematicity is rather nontrivial. In the 2D Heisenberg model, for example, it is viable only in a certain range of parameters such that the magnetic ground state is determined by 
the minimization of fluctuations, which makes it possible the emergence of the spin-nematic order parameter that further can survive above the antiferromagnetic transition temperature \cite{chandra90}. Electron nematicity can also arise from an unequal population of orbitals with different symmetries due to spin-orbital physics \cite{orbital}.
In any case, the absence of direct coupling with external fields makes this scenario quite elusive to experimental verification. Nevertheless, the belief in spin-nematicity has taken roots in the high-$T_c$ community and has motivated a flurry of studies on the iron-based systems. So far most of the attempts have targeted its manifestations in the electronic or magnetic properties with different degree of finality \cite{nematic-on-electrons}.

If the spin-nematic scenario is appropriate to describe the lattice distortion in ferropnictides, the nematic order has to manifest in other properties of the lattice as well. The strong response of the lattice to superconductivity and its softening observed in ultrasound experiments, for example, have recently been interpreted according to this picture in Refs. \cite{nandi10} and \cite{fernandes10} respectively (see also \cite{Kim11} for a structural study). This interpretation, however, does not completely rule out the more standard scenario in which the lattice softens by itself. In this Letter, we indicate an experimental situation in which the spin-nematic mechanism for the structural distortion can be clearly distinguished from a conventional lattice instability. In particular, we show that in epitaxial thin films the spin-nematic mechanism is expected to have a high impact on the structural domains below a certain film thickness.

In epitaxial thin films, the interface with the substrate generates stresses that have been proven effective to increase the superconducting transition temperature in FeTe \cite{han10}. These stresses are known to be responsible for the appearance of different structural domains, whose size is expected to follow the Kittel $l^{1/2}$ law as a function of the film thickness $l$ \cite{roytburd76}. This behavior in fact has been confirmed experimentally in conventional ferroelastics over several orders of magnitude in film thickness \cite{ferroelastics}. The first exception to this rule has been recently reported for the 90$^\circ$ domains observed in PbTiO$_3$/DyScO$_3$ \cite{noheda11}. In this case, lattice misfit plays a key role and deviations from the Kittel $l^{1/2}$ law were predicted long ago \cite{pompe93}. In our case, however, lattice misfit does not break the initial symmetry and therefore is expected to be unimportant as discussed in \cite{bratkovsky01}. Thus, within a pure elastic scenario as described in \cite{barzykin09,cano10}, the Kittel $l^{1/2}$ law is expected to be valid for the tetragonal $\leftrightarrow$ monoclinic instability in the iron-based superconductors. In the following we show that, in contrast, if the spin-nematic ordering is behind this transition, this law is not obeyed in the ultrathin limit. Instead, the size of the monoclinic domains becomes $\sim l_*^2/l$ where $l_*$ is determined by spin-nematic parameters (see below). 
This quantity $l_*$ also defines the corresponding cross-over thickness, which is experimentally accessible as we show below.

The reason for the appearance of ferroelastic domains is the minimization of the stresses generated at the interfaces. This can be described by the Navier-Cauchy (or elastostatic) equations
$\partial_j \sigma_{ij}=0$ in the absence of external forces, where $\sigma_{ij}$ is the stress tensor \cite{landau-elasticity}. For an isotropic elastic medium this tensor is given by the constitutive equation $\sigma_{ij} = \lambda \delta_{ij} \varepsilon_{kk} + 2\mu \varepsilon_{ij}$, where $\varepsilon_{ij}$ is the strain tensor and $\lambda $ and $\mu $ are Lam\' e parameters. The strain tensor, in its turn, can be expressed in terms of the displacement vector $\mathbf u$ as $\varepsilon_{ij} = {1\over 2} (\partial_j u_i + \partial_i u_j )$, what will be convenient in our further analysis. Such a constitutive equation can be obtained as $\sigma_{ij}= {\delta \int F_E dV \over \delta \varepsilon_{ij}}$ from the elastic free energy
\begin{align}
F_E = 
{\lambda \over 2} \varepsilon_{kk}^2  + \mu \varepsilon _{ij}^2. 
\label{FE}\end{align}

Within a Ginzburg-Landau approach, the essential ingredient to describe the nematic-driven structural transition is the free energy
\begin{align}
F = F_E + F_\text{nem} + F_{ME},
\label{F}\end{align}
where 
\begin{align}
F_\text{nem} = 
{a \over 2} \eta^2 + {b\over 4}\eta^4 +
{c_{ij}\over 2}(\partial_i \eta) (\partial_j \eta) 
\label{Fnem}\end{align}
describes the spontaneous appearance of the nematic order parameter $\eta$ and 
\begin{align}
F_{ME}= g 
\left(\varepsilon_{xy} + \varepsilon_{yx}   \right) \eta 
\label{FME}\end{align}
the magnetoelastic coupling that eventually links the nematic ordering with the lattice distortion \cite{note}. The coefficient $a$ is assumed to vary with temperature as $a = a'(T-T_n^{\circ})$ while the rest of parameters are taken as positive constants. Microscopically, $\eta$ can be understood as the result of non-zero long-range correlations between the magnetic degrees of the Fe sublattices, say ${\mathbf m}_1$ and ${\mathbf m}_2$, while $\langle {\mathbf m}_1 \rangle = \langle {\mathbf m}_2 \rangle =0$. 
The establishment of spin nematicity is expected to be a 2D phenomenon taking place within the Fe planes with a rather weak interplane coupling. According to this, the gradient term in Eq. \eqref{Fnem} can be simplified to $c \left[ (\partial_x \eta)^2 + (\partial_y \eta)^2 \right]$. Here and hereafter we assume that the $xy$-plane of the film is parallel to the Fe planes and we refer the axes to the crystallographic unit cell (with more than one Fe atom) where the structural distortion is monoclinic. 

As we have mentioned, the presence of the substrate makes the appearance of nematicity not necessarily uniform in the film, which automatically implies the possibility of having different elastic domains since these follow the nematic ones. In fact, the nematic state has to be compatible with the absence of normal forces at free interface of the film and the vanishing of the displacements deep inside the substrate. That is, with the boundary conditions $\sigma_{xz} (z = l) = \sigma_{yz} (z = l)=0$ and $\mathbf u \underset{z \to -\infty}{=}0$. In consequence the degree of nematicity is expected to vary with the distance $z$ to the interface with the substrate, and consequently in the $xy$ plane to satisfy the equations of equilibrium. We assume that the film length and width are both much larger than any other characteristic length in the problem. Thus, the variation of nematicity in the $xy$ plane is expected to be directly connected to the periodicity of the elastic domains. For the sake of simplicity we assume that the substrate is described by Eq. \eqref{FE} with the same Lam\' e parameters as the film, which does not change qualitatively the results \cite{note1}. We then have elastostatic equations such that the nematic order parameter can be sought in the form $\eta = \eta_0 \cos(\alpha k(z-l))e^{iky}$, which corresponds to a pattern of displacements $\mathbf u = u_0 \cos(\alpha k(z-l))e^{iky}\hat {\mathbf x}$ in the film and $\mathbf u = u_s e^{kz}e^{iky}\hat {\mathbf x}$ in the substrate as can be seen by direct substitution into the linearized equations of equilibrium \cite{note2}. This in fact gives a non-trivial solution of these equations when the temperature is 
\begin{align}
T = T_n^{\circ}+ \Delta T_n^{\circ}\left({1\over 1 + \alpha^2} -  (1 + \alpha^2)(kl_*)^2\right),  
\label{a}\end{align}
where $\Delta T_n^{\circ}= g^2 /(a' \mu) $ and $l_* = (\mu c)^{1/2}/|g|$.
Besides, the parameters $k$ and $\alpha$ have to be such that 
\begin{align}
\alpha = \cot (\alpha k l)
\label{alpha}\end{align}
to ensure that the displacements and the normal forces described by this solution are both continuous at $z =0$, which represents the matching conditions associated to the film/substrate interface. The actual temperature at which the whole system gets unstable with respect to the nematic phase corresponds to the 
highest temperature at which the above two conditions are satisfied, $T_n$ in the following. As in standard second-order transitions, the system is stabilized by the fourth-order term in the free energy \eqref{F} and the amplitude of the order parameter increases continuously as $\eta_0 \sim |T-T_n|^{1/2}$ below $T_n$ (as $u_0$ and $u_s$ do as well). 
This gives a nematic ground state that is obviously degenerate with the solution obtained by replacing $x \leftrightarrow y$ due to the symmetry of the problem. 

\begin{figure}[tb]
\centering\includegraphics[width=.425\textwidth]{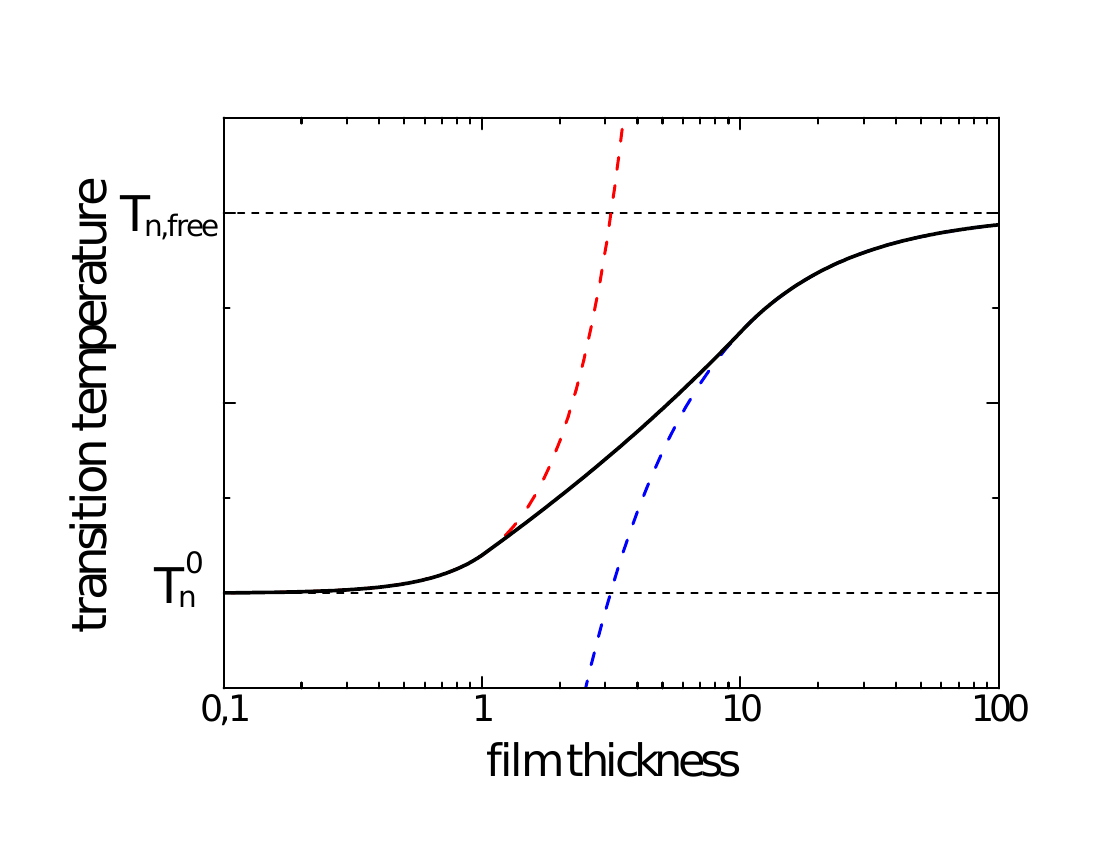}
\caption{Semi-log plot of the transition temperature vs. film thickness (in units of $l_*$) illustrating the expected behavior according to the spin-nematic scenario (black solid line). The blue-dashed line represents the standard behavior of conventional ferroelastics while the red one indicates a $l^2$ dependence. Note that the transition temperature driven by the spin-nematic order is restricted to the limits defined by the nominal transition temperature $T^\circ_n$ and the transition temperature $T_{n,\text{free}}$ of the free standing case. }
\label{Tc}
\end{figure}

The transition temperature as a function of the film thickness is shown in Fig. \ref{Tc}. 
For large values of the film thickness it naturally has an asymptote that corresponds to the transition temperature in the free standing case $T_{n,\text{free}} = T_n^{\circ} + \Delta T_n^{\circ}$. In this limit, it can be said that the dimensions of the film are large enough to achieve a full relaxation of the stresses generated by the interface. The finite dimensions of the film in fact results in a decrease of the transition temperature which initially goes as $T_n \approx (1 - \pi {l_*/l}) T_{n,\text{free}} $. In the opposite case of small thicknesses the transition temperature tends to its nominal value $T_n^{\circ}$. Here it goes as $T_n \approx T_{n}^{\circ}+ \Theta(l/l_*)^2$ 
in the ultrathin limit, where $\Theta = c /(4 a' l_*^2)$. The crossover between these two regimes is determined by the quantity $l_*$.

\begin{figure}[tb]
\centering\includegraphics[width=.4\textwidth]{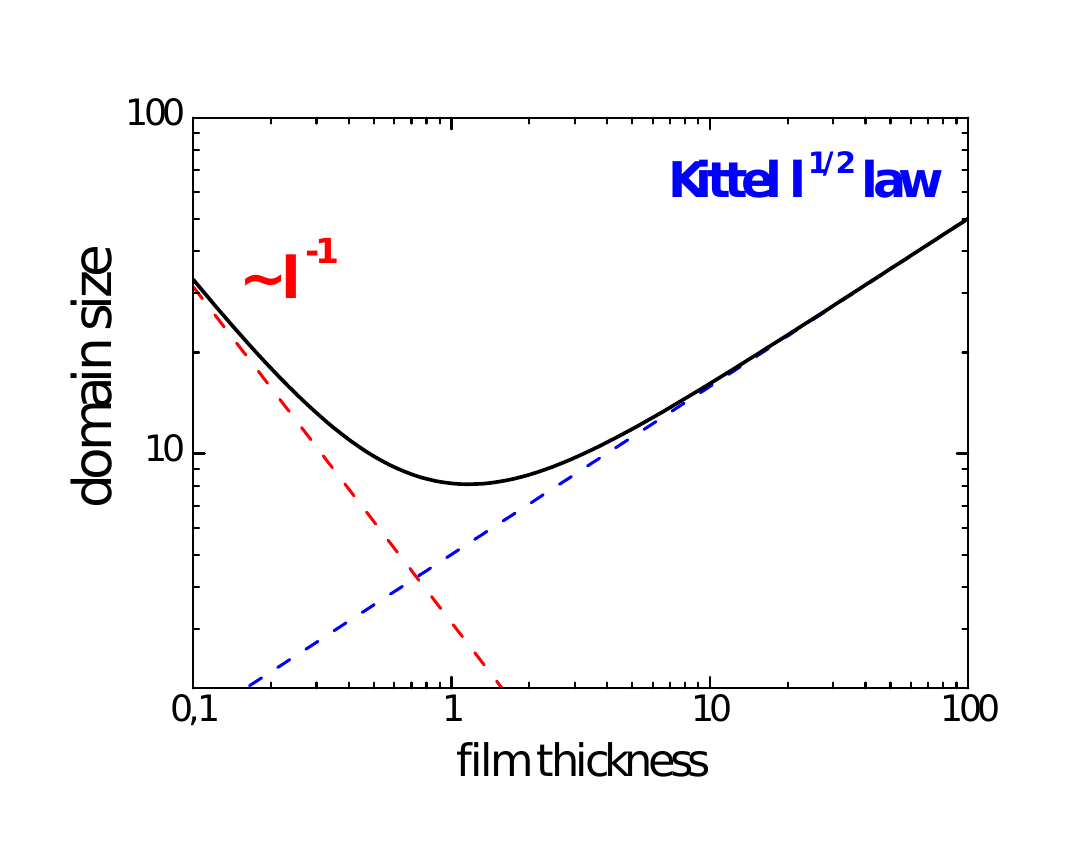}
\caption{Log-log plot of the domain size vs. film thickness in units of $l_*$ that illustrates the expected cross-over from the Kittel regime to a new quasi-single domain regime for the ultrathin limit.}
\label{size}
\end{figure}

Fig. \ref{size} displays the size of the elastic domains as a function of the film thickness. For thicknesses $l \gg l_*$ the Kittel $l^{1/2}$ law is obeyed. In this regime the size of the domains is however $\sim (8\pi)^{1/2}( l_* l )^{1/2}$, so it already reflects the magnetoelastic coupling through the quantity $l_*$. In fact the size of the domains turns out to be similar to the film thickness when this is $l \sim l_*$, and the same happens with the region in which the deformation of the substrate is concentrated. Below this point, the system starts to deviate from the Kittel behavior. The reason is that the formation of new domains does not help to minimize the stresses anymore. In fact, in the ultrathin limit ($l \ll l_*$), these deviations are very marked and the domain size unusually increases as $\sim \pi l_*^2/l$ with decreasing film thickness. This can be understood as due to the energy penalty associated with the gradients in the corresponding domain walls. 

These results are based on a mean-field treatment of the nematic transition that is expected to be valid even in the ultrathin limit. The reason is that the nematic excitations remain gapped at the critical point except along two soft lines in $\mathbf k$ space, which is due to the magnetoelastic coupling \eqref{FME}. This increases the effective dimensionality of the problem and makes critical fluctuations eventually irrelevant \cite{cano10} (see also \cite{qi09
}). 

As we see, the quantity $l_*$ defines a crossover between two different regimes in which the system behaves very differently. For $l \gg l_*$, on one hand, the system effectively behaves as a conventionial ferroelastic. That is, there is a change in the transition temperature inversely proportional to the film thickness and the ferroelastic domains that appear at this point follow the Kittel $l^{1/2}$ law. On the other hand, for $l\ll l_*$, the fact that the actual instability is driven by the nematic order parameter shows up very clearly. In this regime i) the transition temperature tends to its nominal value (in contrast to a conventional ferroelastics where it is expected to decrease) and ii) the system tends to get rid the domain walls and the size of the domains increases by decreasing the film thickness. The observation of this crossover in iron-based superconductors would be a direct probe that the spin-nematic scenario is taking place.

The precise value of $l_*$ is quite difficult to anticipate due to the absence of direct measurements of the bare parameters that determine this quantity. Nevertheless, its order of magnitude can be estimated as follows. Since the spontaneous strains are assumed to be induced by the nematic order parameter, no special softness of the lattice itself has to be invoked. $\mu$ then can be roughly estimated as the atomic energy $E_\text{at}\sim \rm eV$ per unit cell. Experimentally, the spontaneous strains $\varepsilon = -g \eta /\mu$ saturate to $\sim 10^{-3}$ in bulk samples. In units such that this corresponds to $\eta = 1$, this gives the ratio $|g|/\mu \sim 10^{-3}$. The coefficient $c$, in its turn, can be estimated as $c \sim E_n l_\text{at}^2 $ taking into account that space variations of the order parameter from $\eta =1$ to $-1$ within the interatomic distance $l_\text{at}$ ($\sim 4${\AA}) are expected to be associated with the characteristic energy $E_n \sim \rm meV$ of the nematic order. This gives $l_* \sim 10^3 (E_n /E_\text{at})^{1/2}l_\text{at} \sim 12 \rm nm$, which corresponds to $~10-15$ times the typical lattice parameter along the $c$ direction. This conservative estimate indicates that the cross-over thickness $l_*$ is accessible experimentally.

We finally make the rather obvious remark that, by means of the structural domains that are expected to appear in epitaxial films, one can extract valuable information about the hypothesized spin-nematic state. At the mean-field level this state is described by Ginzburg-Landau free energy \eqref{F}, which in principle can be derived from more microscopic approaches as the worked out in \cite{MM}. In Eq. \eqref{F} we have the parameter $T_n^\circ$ describing the nominal transition temperature that, as we have discussed, corresponds to the limiting value of the transition temperature that will be observed experimentally for the thinnest samples. Furthermore, the parameters $a'$, $c$ and $g$ are associated to the quantities $\Delta T_n^\circ$, $l_*$ and $\Theta$ in such a fashion that can all be extracted from the appropriate experimental data. The information obtained in this way can further be contrasted with other independent measurements in order to get a phenomenological characterization of the spin-nematic ordering.

In summary, we have shown that the spin-nematic mechanism proposed for the structural transition in iron-based superconductors should reveal itself in the form of strong deviations from the Kittel behavior exhibited by standard ferroelastics. We have predicted an intrinsic cross-over thickness $l_*$ below which a quasi-single domain regime should emerge. In this regime, the size of the domains is expected to be inversely proportional to the film thickness at the same time that the transition temperature saturates to its nominal value. We have indicated the viability of this cross-over which is connected to the magnetoelastic coupling in the ferropictnides. 
These findings are expected to open new routes for the study the spin-nematic physics, and are relevant for pseudo-proper ferroelastics in general. 

I acknowledge E. Bascones, J.E. Lorenzo, D. Mannix, M. Nu\~ nez-Regeiro, I. Paul and especially E. Kats for stimulating discussions.


\begin{references}
\bibitem{note1} The difference between $\mu$ in the film and $\mu'$ in the substrate, which in any case is expected to be small ($\mu/\mu ' \sim 1$), modifies Eq. \eqref{alpha} as $\alpha = (\mu / \mu') \cot (\alpha kl)$. This further changes the size of the domains simply as $\sim (\mu/\mu')^2(l_*^2/l)$ in the ultrathin limit ($l\ll l_*$).
\bibitem{note2} The periodicity of the solution and its degeneracy with respect to its absolute phase are due to the initial translational symmetry of the system in the $xy$ plane.
\end{references}
\end{document}